# Development of simulation model for Single Carrier Transciver for Nanosatelite

Pallewela R.C.K
Curtin university of Technology

Dr. Rohana Thilakuamra
*Main Supervisor*

Mr.Prabath Buddhika
Co-Supervisor

***Abstract*— Cubsat (CubeSat) is a nanosatellite concept emerged from a paper published by Stanford University [1] and with their low cost nature and extreme feasibility , more started researching on nano satellites. [2] New technology emerged , paving path to many academics and small vendors to create their own Cubesat models .**

**This nanosatellite requires a transceiver to maintain its communication between it's systems and the ground station, which helps it navigate and collects data gained from its programmed functions. This transceiver system consists mainly of a transmitter and a receiver. The transmitter manages sending data from the satellite to ground station while the receiver captures the data and instruction sent from the ground station to the satellite.[7]**

**These systems were built using separate digital communication devices in the beginning, with many critical limitations with respect to the space and scalability of the modules to be attached and the programmability of hardware materials and the concept of system-on -board emerged. Meanwhile, As the size of electronic devices minimized with the research conducted, FPGA (Filed Programmable Logic Array) was introduced as an architecture to be used in various applications and research needs [23]. The reason FPGA was mention was for the fact that it provides flexibility in the designing of transceiver ed with design and prototypes implementation at a low cost comparitional electronic ware and the system -on chip Concept was introduced . This research describes the development a system-on-chip transceiver model for nanosatellites which contains a single carrier .**

## I. INTRODUCTION

The growth of interest in the Satellite communication had to lead the development of more efficient and budget friendly satellite types like Nano satellite. CubeSat is one of the nanosatellite concepts emerged from a paper published by Stanford University [1]. CubeSat's sub systems are all located inside a cube (10 x 10 x 10 cm3), taking storage constrains into account. Compared to the traditional communication satellites, the CubeSat's nature of short development time, less expense and ability to perform the same tasks as of traditional ones for a shorter duration has pushed more project developers to be conducting projects by accepting the higher risks and the mission failure probability and thus, new developing ear began for the Cubesats. [2]. From then to today, a rapid increase in technology has occurred, introducing the system on-board communication subsystem for Cubesat. [3] The communication subsystem of a Cubesat consists of transceiver, transponder, and the antennas and according to the system on board concept they all were embedded into a single board. This cleared up more space for more powerful battery and solar panel to be installed to maintain the satellite for a longer time in space. As the need for storage space increased, concepts on system on chip emerged. [4] Satellite development companies has come up with different models for this concept but due to the high demand in market, the system development stages are not disclosed to other industrial or academic purposes. The CubeSat, although less expensive than of the traditional satellites, it still is out of the budget of for most of the small-scale research developers. Thus, a model using an open-source hardware header in C++ called SystemC, which can easily be translated into Hardware description language (HDL) for the purpose of development of "receiver", a component of communication subsystem will be introduced. And the validity of the model will be confirmed by converting and executing selected subsystems of the said receiver in Verilog FPGA.

## II. BACKGROUNG READING

The communication sub system in a satellite's main task is to maintain good communication between the ground station ; where execution instructions are send from earth to be executed at the OBC.

In idle state, the nanosatellite is on stand by mode till a signal is send and then the cube sat immediately sends an acknowledgement signal (ACK) through the transponder mentioning the signal was received. Then the Cubesat gradually extend the antennas and gather requested data or transmit the gathered data to the ground station using the transmitter.

For this communication to happen, according to the [5] paper, There is a communication window present in order to establish communication.

This is tested and is noted in scheduler, so the transmission of data can be done in that time period.

i. Transponder – Immediate responder for a transmitted wireless signal

ii. Transceiver _ the transmission and receiving if data is done by this module

iii. Antennas- Used to grasp the data send to the receiver and to transmit data in correct angel with line of sight communication managed with the ground station.

### *A. Transciver Model*

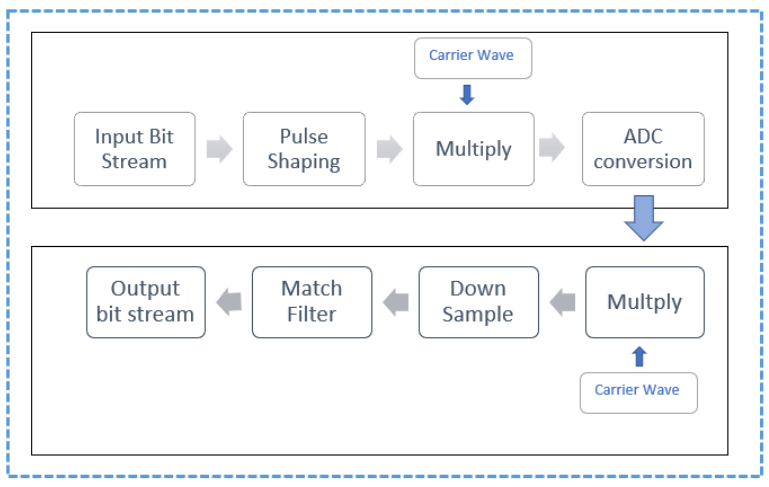

*Figure 1 Transceiver Model*

- Input Bit stream – This is the signal send in to the system
- Pulse shaping - The band width of the bits to be transmitted can be band limited using this technique
- Line coding- Before the pulse shaping is performed, the input bit stream should be translated to a wave form. There are many line coding techniques available.[6]
- Carier wave – This wave modulates information baring signals in to a the transmit wave form . The frequency is much higher than of the information signal wave.
- Modulation –Modulation can be of two parts depending on the analog and the digital nature

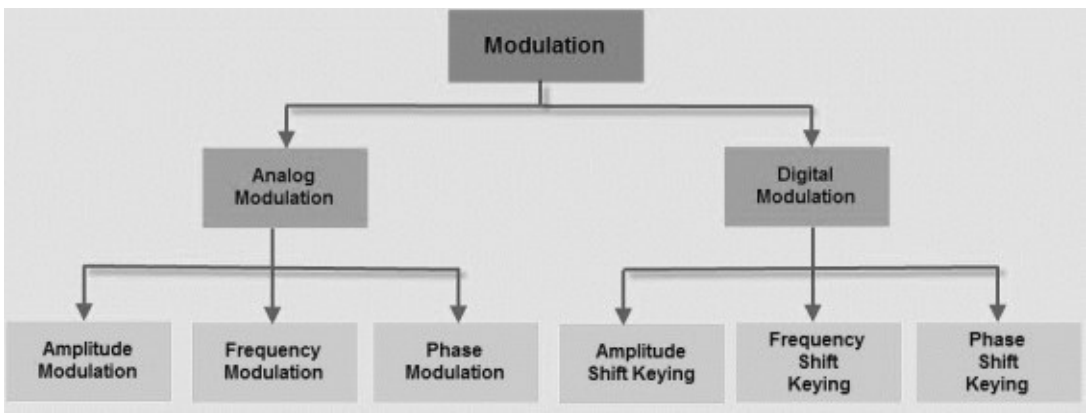

*Figure 2 Modulation Techniques*

- In nano satellite communication the most commonly used modulation technique is AFSK (Audio Frequency Shift Keying )but in [7] it is proven that BPSK ( Binary Phase Shift Keying ) is a better choice due to the Bit error Rate (BER) .

| Modulation scheme | EB/N0 for BER 10-5 | Constant envelope? | Spectral efficiency |
|---|---|---|---|
| AFSK | 23dB | Y | Bad |
| BPSK | 9.6dB | Y | Bad |
| Shaped BPSK | 9.6dB | N | Good |

*Figure 3 Modulation for Nanosatellites [7]*

- Bit error rate -The Ratio between the received bits and transferred bits .
- ADC Conversion – The analog data being converted to discrete data by sampling and quantizing [8]
- Demodulation – The transferred signal is multiplied with the same frequency carrier which is locally generated in order to obtain the required data [8]
- Match filter – The signal is then passed thought a low pass filter in order to remove impulses with in signal and to band limit it .
- By considering the above receiver, the conclusion to that the receiving signal error can affect the output signal is noted.
- To overcome and minimize this error, a phase locked loop receiver or a Costas loop receiver can be utilized

### B. *Costa Loop ( Phase Locked Loop -PLL)*

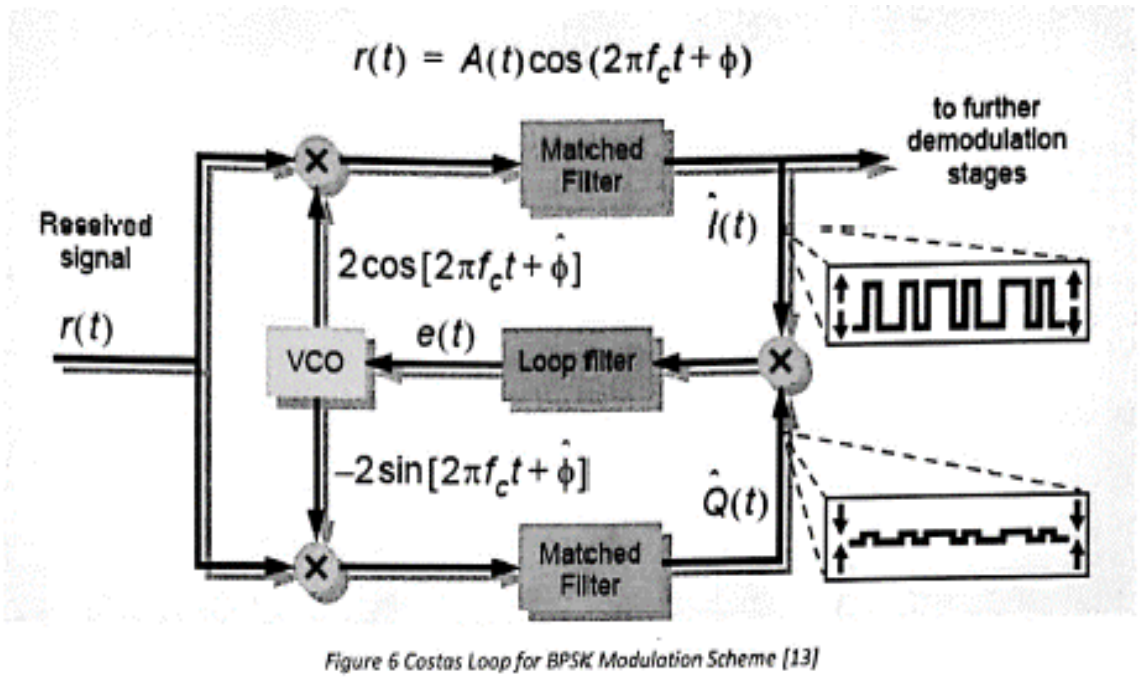

*Figure 4 Costas Loop [8]*

The received signal is sent through two different carrier signal, but with the same frequency and then a matched filter (Low pass filter) is used to filter data in orderly manner. The loop is controlled by a VCO (Voltage Controlled Oscillator) and the Direct Current (DC) level changes according to the VCO to recover the frequency and carrier of the signal

## III. METHODOLOGY

The Purpose of this project is to model a single carrier transceiver in C++ (using SystemC header) and to implement a few sub systems in FPGA depending on availability of time.

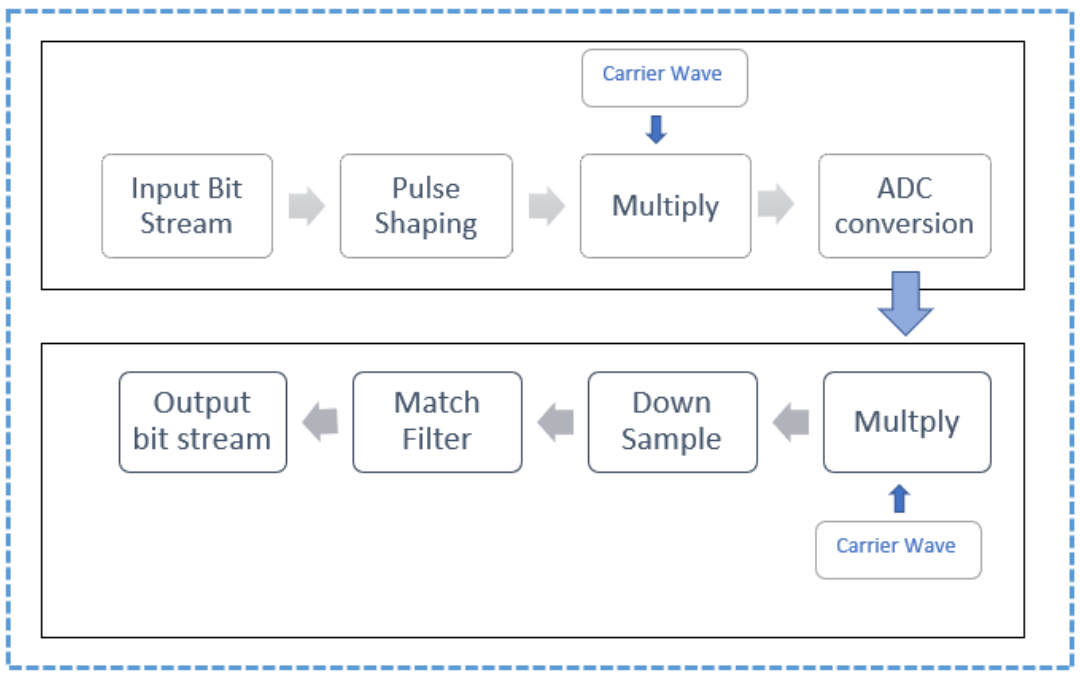

*Figure 5 Purposed Model*

### A. *Transmitter Hardware Model*

After Going through The SystemC datasheet and the overall diagram of transmitter , it was decided that the a 2x1 Mux model can be used in modelling the transmitter

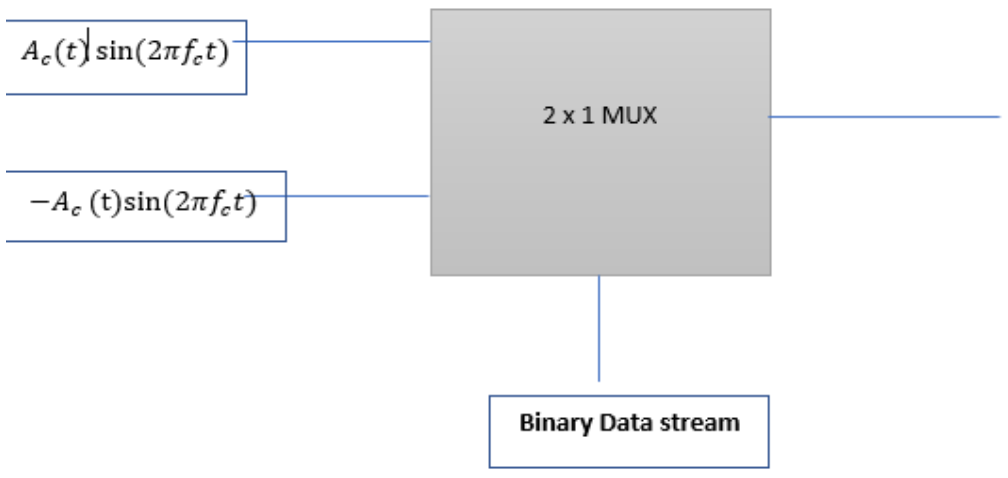

*Figure 6 Modeling Transmitter*

## B. Reciver Hardware Model

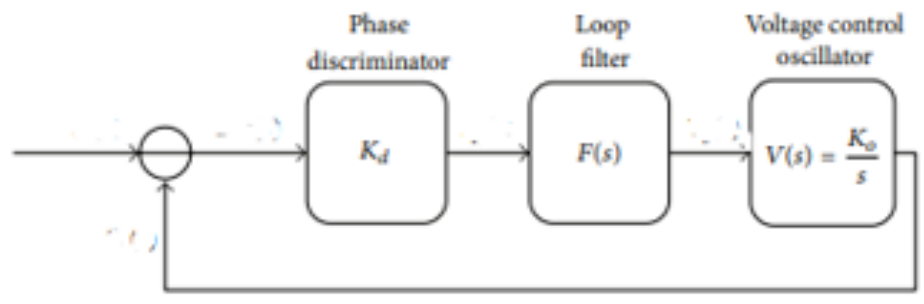

*Figure 7Modeling Receiver[9]*

The Phase discriminator was modeled using a Flipflop circuit

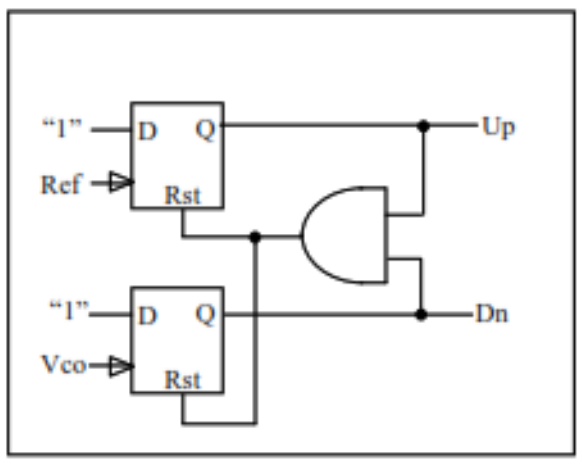

*Figure 8 Modeling Phase Discriminator [9]*

The loop filter was designed as a shift register.

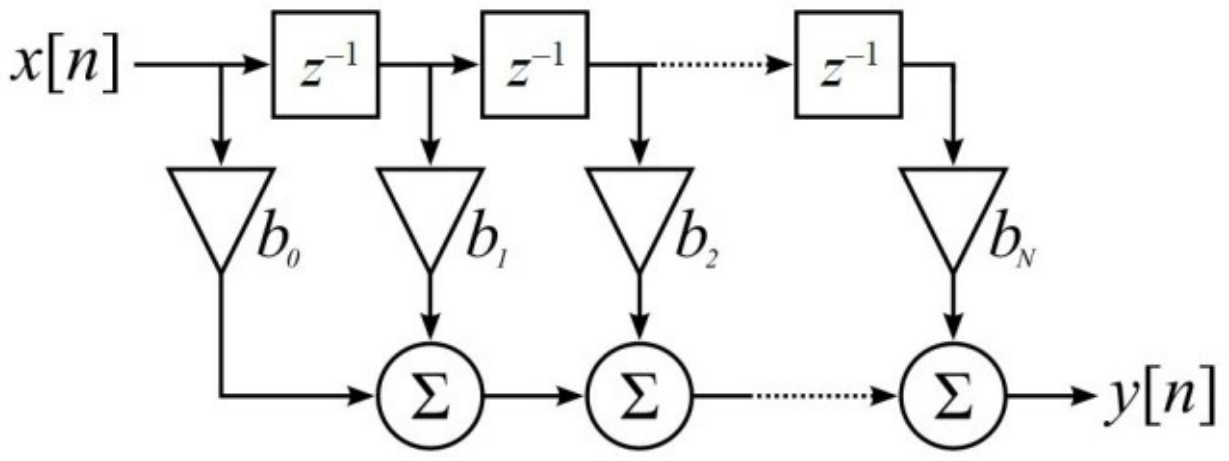

*Figure 9 Shift register [9]*

The VCO ( Voltage Controlled Oscillator) was designed using a MUX (Multiplexer Circuit)

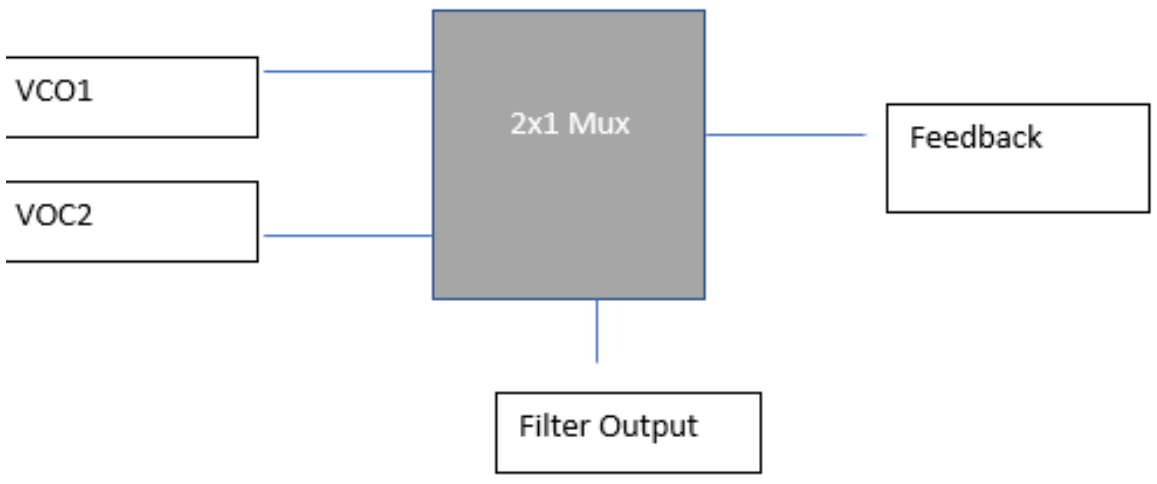

*Figure 10 VCO modelling*

# IV. IMPLEMENTATION AND RESULTS

## A. SystemC/C++ model

- The systemC/C ++ module gave the following out come.

a. The NRZ (no return to zero ) unipolar sequence was send through the modulator

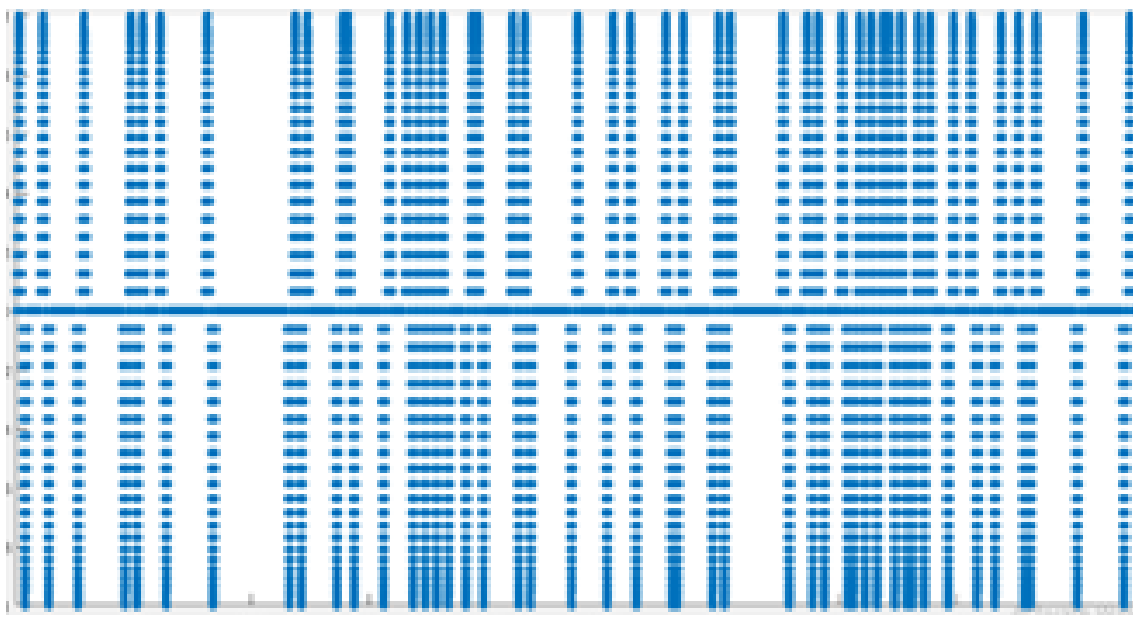

*Figure 11BPSK modulated signal*

## B. Verilog /ModelSim Output

The DSS( Direct Digital sequence Implementation

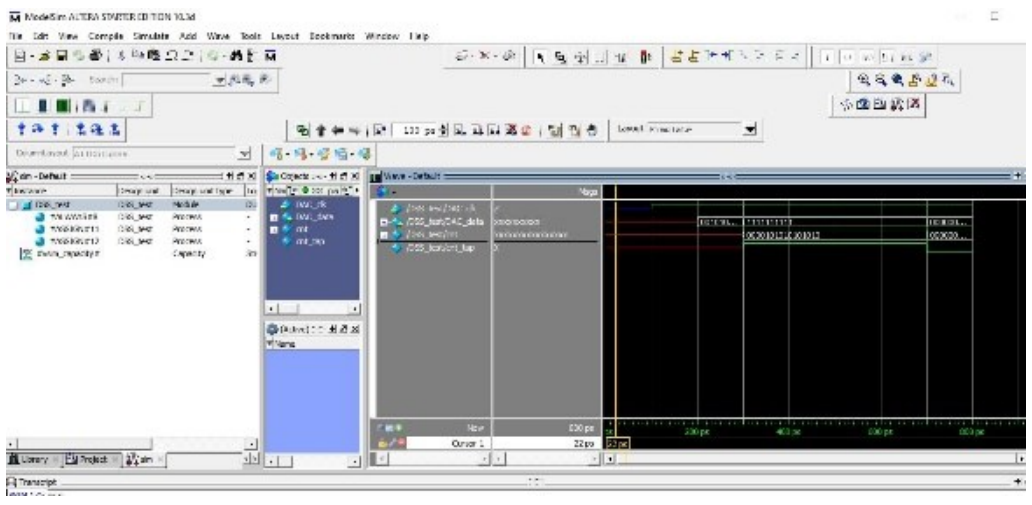

*Figure 12The DSS( Direct Digital sequence) to input sin data to MUX*

The MUX( 2:1 Multiplexer) Implementation

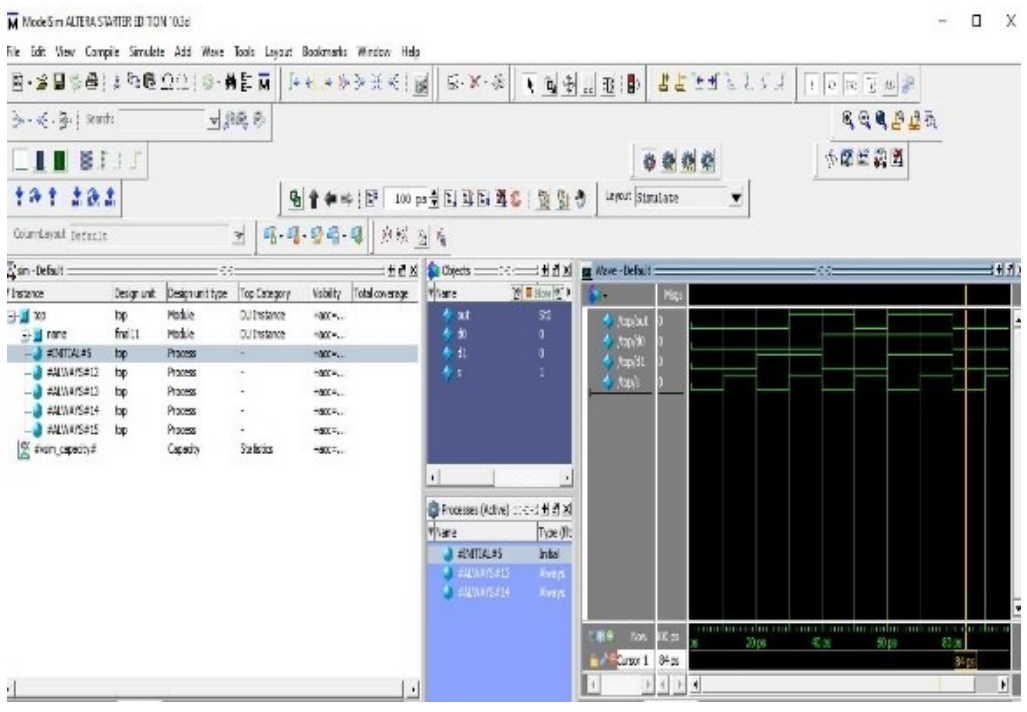

*Figure 13MUX( 2:1 Multiplexer)*

The Low pass Filter (LPF) implementation

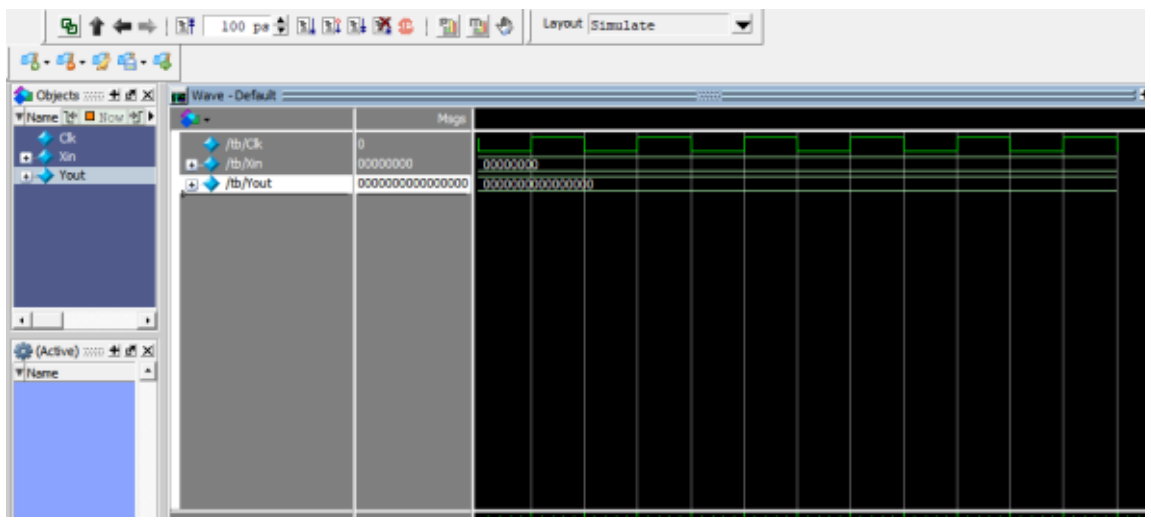

*Figure 14Low pass Filter (LPF) implementation*

## V. Conclusion

The objective of this project was to develop a working simulation model for transceiver in an open source platform and from the analysis the following points were made. The simulation process was a success but the limitation of systemC language deviated us from developing a systemC generic model to a c++ and systemC model, in order to send analog signals for modulation. The Verilog requirement was fulfilled by implementing 3 modules .

## Acknowledgment

I would like to extent my gratitude to Dr. Rohana Thilakumara for his guidance and patience during the course of the project. The theoretical doubt clarification and the time and advice lent to me and encouraging me to work on the project efficiently was remarkable. Next, I would like to extend my sincere thankfulness to Mr. Prabath Buddhika , my co-supervisor for his support throughout the project. In addition, I would also like to thank Ms. Sachini Kandawala ,my unit coordinator who was very patient and understanding about the whole course of time. Finally , I would like to extend mt gratitude to my parents for their unwavering support in my life . The encouragement, moral and monetary support given by them to complete this project was the reason it was possible for me to finish this project successfully.